\documentclass[twocolumn,showpacs,preprintnumbers,amsmath,amssymb]{revtex4}

\usepackage{graphicx}
\usepackage{dcolumn}
\usepackage{bm}

\begin{document}

\title{Improving imaging resolution of shaking targets by Fourier-transform ghost diffraction}

\author{Cong Zhang}
\author{Wenlin Gong}
\email{gongwl@siom.ac.cn}
\author{Shensheng Han}
\affiliation{ Key Laboratory for
Quantum Optics and Center for Cold Atom Physics of CAS, Shanghai
Institute of Optics and Fine Mechanics, Chinese Academy of Sciences,
Shanghai 201800, China}

\date{\today}

\begin{abstract}
For conventional imaging, shaking of the imaging system or the
target leads to the degradation of imaging resolution. In this
work, the influence of the target's shaking to fourier-transform ghost
diffraction (FGD) is investigated. The analytical results, which are
backed up by numerical simulation and experiments, demonstrate
that the quiver of target has no effect on the resolution of
FGD, thus the target's imaging with high spatial resolution can be always achieved
by phase-retrieval method from the FGD patterns. This approach can be applied
in high-precision imaging systems, to overcome the influence of the system's shaking
to imaging resolution.
\end{abstract}


\maketitle

Relative motion between the imaging system and the object is one of the common reasons for
imaging degradation. For instance, camera's shaking or object's high-speed moving would cause the blur
and degradation of imaging resolution. For conventional imaging, a usual
method to decrease motion blur is to reduce the exposure time, while
this will reduce the signal-to-noise ratio (SNR) of the images.

Other approaches for motion deblurring include mechanical compensation, optical
compensation and electrical compensation \cite{Olson,Lareau}. However, the imaging resolution is limited
by the compensation accuracy. Also, the designing and manufacturing of compensation devices are complex
and the development of compensation techniques is still one of the most challenging tasks.

Motion deblurring can also be completed by digital image processing
method \cite{Gonzalez}. The blur kernel, namely point spread function, is the key factor in
the process of image restoration. If the blur kernel is known before image restoration, inverse filtering and
Wiener filtering are widely used for image restoration \cite{Wiener,Lagendijk,Tan}. If the blur kernel cannot be known or measured in some
practical applications, image restoration is usually realized by blind deconvolution technique \cite{Lane,Lane1,Rav-Acha}. However,
image restoration needs large computation and suffers from serious restoration errors, moreover it is sensitive to noise disturbance.

Ghost imaging (GI), as a nonlocal imaging method, has been widely investigated
during last fifteen years \cite{Strekalov,Ribeiro,Bennink,Zhang1,Gatti1,Gatti2,Angelo,Cheng}.
An interesting and obvious feature of ghost imaging is that fourier-transform
imaging of the object can be obtained by measuring the intensity correlation function
between two light fields, even when the source is a thermal light source and the
object is located in Fresnel region relative to the source \cite{Cheng,Zhai,Xiong,Zhang2,Ying,Wang}.
Recently, super-resolution far-field ghost imaging and ghost imaging lidar are also obtained
by exploiting the sparsity constraints of images \cite{Gong1,Gong2,Zhao}. However, all the previous work on
ghost imaging is focused on the static targets. Obviously, imaging an unstable and moving target is meaningful in practice.
Based on the lensless fourier-transform ghost diffraction (FGD) scheme \cite{Zhang2} and phase-retrieval techniques \cite{Ying},
the effect of the target's shaking on FGD is investigated both theoretically and experimentally, and the target's imaging is retrieved
from the experimental FGD patterns. We show that the scheme is helpful in overcoming the shaking blur due to the novel characteristic of ghost imaging technique.

\begin{figure}[htbp]
\centering
\includegraphics[width=8.5cm]{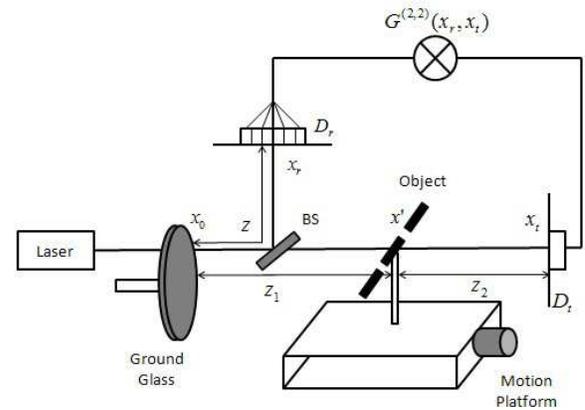}
\caption{Schematic of lensless FGD for shaking targets with
pseudo-thermal light. The object is operated by a motion platform so
that it can shake on the object plane, perpendicular to the
optical axis.}
\end{figure}

Fig. 1 presents the experimental schematic of lensless
FGD for shaking targets with
pseudo-thermal light. The pseudo-thermal source, which is obtained
by modulating a laser beam (the wavelength $\lambda$=532 nm and the
light spot's transverse size $D$=4 mm) with a rotating ground glass,
is divided by a beam splitter (BS) into a test and a reference
paths. In the reference path, the light propagates directly to a
charge-coupled device (CCD) camera $D_r$. In the test path, the
light goes through an object and then to a single pointlike
detector $D_t$. The object is placed on a motion platform and the
platform is driven by a stepping motor so that it can move straight
in one dimension. In addition, the stepping accuracy of the platform
is 5 $\mu m$ and the stepping motor is connected to a computer,
which can control the motion of the object.

Based on GI theory \cite{Gatti2,Cheng}, the correlation function
between the two detectors is
\begin{widetext}
\begin{eqnarray}
\Delta
G^{(2,2)}(x_r,x_t)&=&\left\langle{E_r(x_r)E_r^*(x_r)E^*_t(x_t)E_t(x_t)}
\right\rangle-\left\langle{E_r(x_r)E_r^*(x_r)}
\right\rangle\left\langle{E_t(x_t)E^*_t(x_t)}
\right\rangle\nonumber\\&=&
\int{dx_{01}dx_{02}dx_{03}dx_{04}}\left\langle{E_0(x_{01})E_0^*(x_{04})}
\right\rangle\left\langle{E_0^*(x_{02})E_0(x_{03})} \right\rangle
\nonumber\\&\times&\left\langle{h_r(x_r,x_{01})h_t^*(x_t,x_{04})h_r^*(x_r,x_{02})h_t(x_t,x_{03})}
\right\rangle,
\end{eqnarray}
\end{widetext}
where $<\cdots>$ denotes an ensemble average, $E_r(x_r)$ and $E_t(x_t)$ denote the light fields on the
reference and test detection planes, respectively. $\left\langle
{E_0(x_{0})E_0^*(x_{0}')} \right\rangle$ is the auto-correlation
function of light field on the ground glass plane. $h_t(x_t,x_0)$
is the impulse response function of the test path, whereas
$h_r^*(x_r,x_0)$ denotes the phase conjugate of the impulse response
function in the reference path.

For the schematic shown in Fig. 1, under the paraxial approximation,
the impulse response function of the reference path is
\begin{eqnarray}
h_r(x_r,x_0)\propto\exp\{ \frac{j k }{2z}(x_r-x_0)^2\},
\end{eqnarray}
where $k=2\pi/\lambda$. And the impulse response function for the
test path is
\begin{eqnarray}
h_t(x_t,x_0)&\propto&\int{dx'} t(x', \tau)\exp\{ \frac{j
k}{2z_1}(x'-x_0)^2\} \nonumber \\ &\times& \exp\{ \frac{j
k}{2z_2}(x_t-x')^2\},
\end{eqnarray}
where $t(x',\tau)$ is the transmission function of the object at time $\tau$. Assuming
that the light field on the ground glass plane is fully spatially
incoherent and the intensity distribution is uniform as a constant intensity $I_0$, then
\begin{eqnarray}
&\left\langle{E_0(x_{01})E_0^*(x_{04})} \right\rangle=I_0 \delta(x_{01}-x_{04}), \nonumber \\
&\left\langle{E_0(x_{02})E_0^*(x_{03})} \right\rangle=I_0
\delta(x_{02}-x_{03}),
\end{eqnarray}
where $\delta (x)$ is the Dirac delta function.

Substituting Eq. (2)-(4) into Eq. (1) and suppose that $z=z_1+z_2$, the
correlation function is
\begin{eqnarray}
\Delta G^{(2,2)}(x_r,x_t) \propto \int {dx_1'}{dx_2'}\exp \{-\frac{jk}{z_2}(x_r-x_t)x_1'\}\nonumber \\
\times \left\langle{t(x_1', \tau)t^*(x_2', \tau)} \right\rangle \exp
\{\frac{jk}{z_2}(x_r-x_t)x_2'\},
\end{eqnarray}

The object's shaking means that the center position of the object
will deviate from the optical axis. If the object's center position
deviates $\xi$ from the optical axis at time $\tau$, then the object's transmission
function $t(x',\tau)$ will change into $t(x'-\xi)$. In practical
applications, although the shaking deviation of the object is unknown and
random at every time, it obeys to a certain statistical distribution.
Assume that the shaking deviation $\xi$ obeys to probability
distribution $P(\xi)$ (where $\int{d\xi} P(\xi)=1$), then
\begin{equation}
\left\langle{t(x_1',\tau)t^*(x_2',\tau)} \right\rangle=\int{d\xi} P(\xi)
t(x_1'-\xi)t^*(x_2'-\xi),
\end{equation}

Substituting Eq. (6) into Eq. (5), after some calculations, Eq. (5) can be represented as
\begin{eqnarray}
\Delta G^{(2,2)}(x_r,x_t) &\propto& \left|\tilde
T(\frac{2\pi}{\lambda z_2}(x_r-x_t))\right|^2 \int{d\xi}P(\xi)
\nonumber \\&=&\left|\tilde T(\frac{2\pi}{\lambda
z_2}(x_r-x_t))\right|^2,
\end{eqnarray}
where $\tilde T(q)$ is Fourier transformation of the object $t(x')$. If the
detector in the test path is a single pointlike detector which is
located at $x_t=0$, then
\begin{equation}
\Delta G^{(2,2)}(x_r,x_t=0) \propto \left|\tilde T(\frac{2\pi}{\lambda
z_2}x_r)\right|^2.
\end{equation}

From Eqs. (7) and (8), the object's FGD can be still obtained even if the object is shaking and 
the result is the same as that achieved in the static condition \cite{Cheng}. Therefore, the
object's shaking has no effect on the imaging resolution of FGD.

In order to verify the analytical results, Fig. 2 and Fig. 3,
respectively, present the numerical simulated and experimental
results of imaging a three-slit in different shaking conditions, using the lensless scheme shown
in Fig. 1. In the simulation and experiments, the distances listed in
Fig. 1 are as follows: $z=550$ mm, $z_1=200$ mm, $z_2=350$ mm. The
object is a three-slit with slit width $a=0.2$
mm, center-to-center separation $d=1.0$ mm, and slit height $h=1.0$
mm.

\begin{figure}[htbp]
\centering
\includegraphics[width=8.5cm]{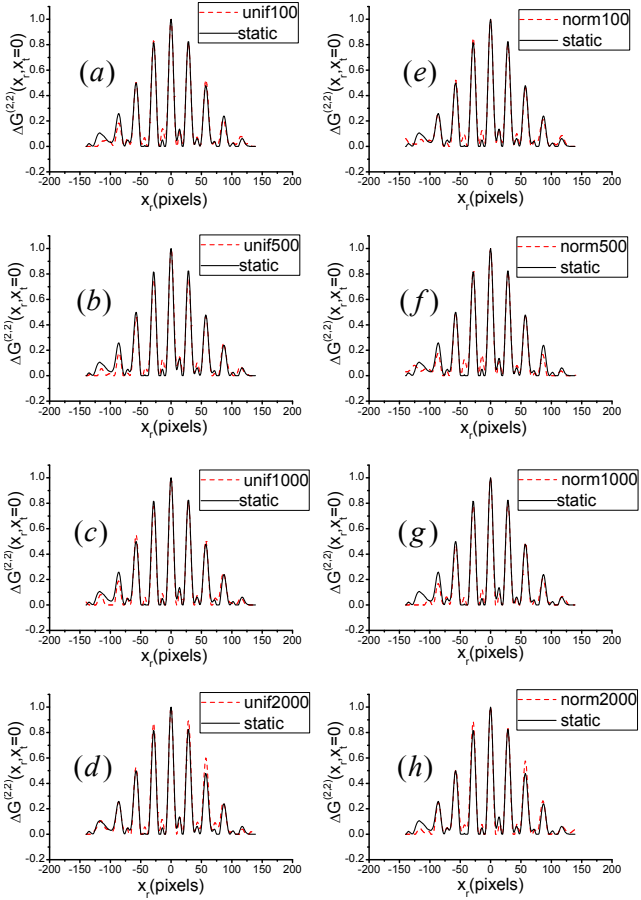}
\caption{Numerical simulation results of shaking object
FGD with the lensless scheme shown in Fig. 1 (averaged 4000 measurements). X-axis indicates the position on the reference detector $D_r$. Y-axis indicates
FGD reconstructed by measuring the intensity correlation function $\Delta G^{(2,2)}(x_r,x_t=0)$.
(a)-(d): the shaking mode obeys to uniform statistical distribution, and the maximum shaking deviation from the optical axis is
$100\mu m$, $500\mu m$, $1000\mu m$, $2000\mu m$, respectively. (e)-(h): the shaking mode obeys to normal statistical distribution, and the maximum shaking deviation from the optical axis is
respectively $100\mu m$, $500\mu m$, $1000\mu m$, $2000\mu m$. Red dashed curves present FGD reconstructed in different shaking conditions versus black solid curves for the static condition.}
\end{figure}

\begin{figure}[htbp]
\centering
\includegraphics[width=8.5cm]{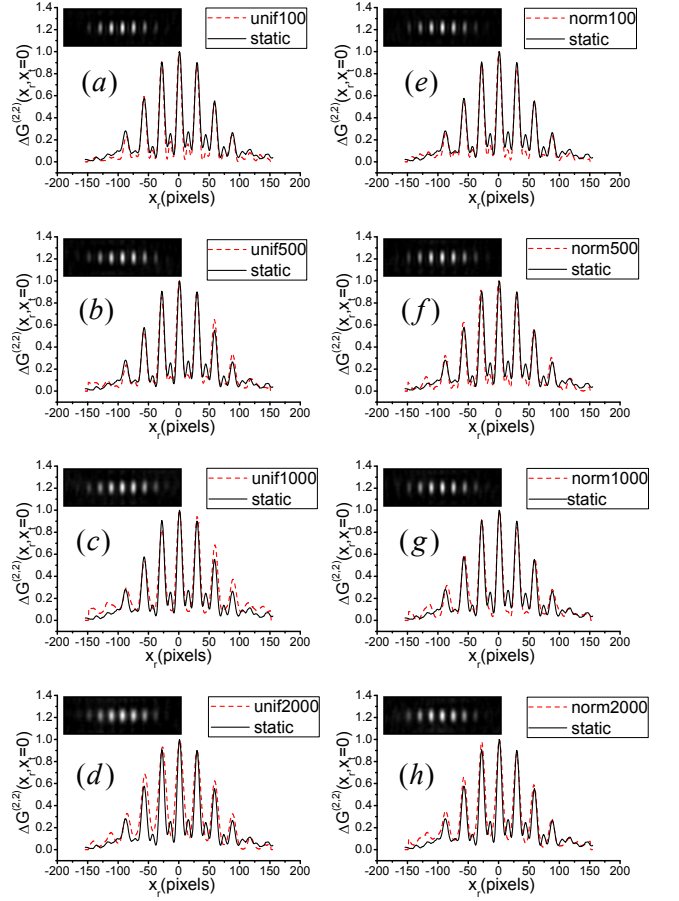}
\caption{Experimental results of shaking object FGD, the conditions are the same as Fig. 2.
The images in different shaking conditions reconstructed by measuring the intensity correlation function $\Delta G^{(2,2)}(x_r,x_t=0)$ \
are listed at the upper left corners and red dashed curves are their cross-sections. In contrast, black solid curves are the cross-section
of FGD obtained in the case of static condition.}
\end{figure}

The shaking modes of the object are controlled by the motion
platform and two statistical distributions are discussed: uniform
and normal. The maximum shaking deviation from the optical axis is
$100\mu m$, $500\mu m$, $1000\mu m$, $2000\mu
m$, respectively, which is used to perform the influence of the
shaking amplitude to the imaging resolution of FGD.

For Fig. 2 and Fig. 3, FGD in the static condition is illustrated by black solid curves. If the object's shaking mode obeys to uniform
distribution, the results in different shaking amplitudes are shown in
Fig. 2(a)-(d) and Fig. 3(a)-(d). While in the case of normal
distribution, Fig. 2(e)-(h) and Fig. 3(e)-(h) present the
corresponding simulated and experimental results. From the simulated
and experimental results shown in Fig. 2 and Fig. 3, both the
shaking mode and the shaking amplitudes of the object have no effect on
the imaging resolution of FGD and the imaging quality is the same as
that in the static case, which is in accordance with the analytical
results described by Eqs. (7) and (8).

\begin{figure}[htbp]
\centering
\includegraphics[width=8.5cm]{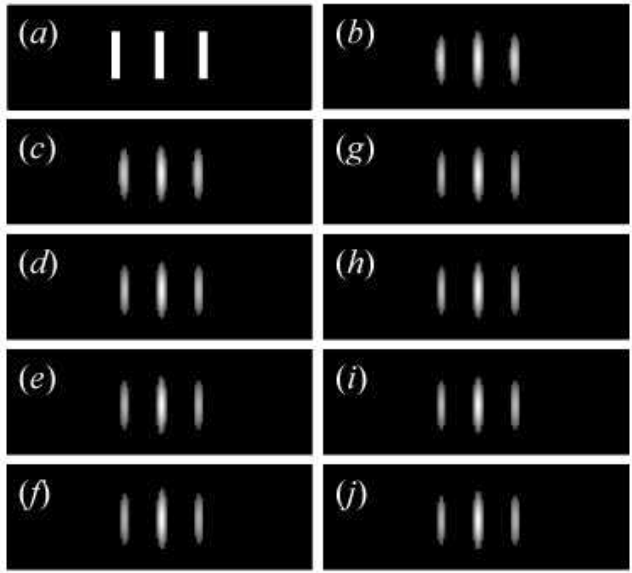}
\caption{The object's imaging retrieved from FGD
patterns based on the experimental results of Fig. 3.
(a) The original object used in numerical simulation and experiments;
(b) phase-retrieval image of static object FGD; (c)-(f): corresponding phase-retrieval results of Fig. 4(a)-(d);
(g)-(j): corresponding phase-retrieval results of Fig. 4(e)-(h).}
\end{figure}

In addition, based on phase-retrieval iterative algorithm \cite{Ying},
the object's imaging in two shaking modes and different shaking amplitudes, as shown in Fig. 4, can be retrieved
from the experimental FGD patterns of Fig. 3. From Fig. 4(d,f) and (i,j), even if the object's shaking amplitude is much larger
than center-to-center separation of the object, the spatial resolution of reconstructed object is nearly the same as that retrieved in the static condition
when FGD technique and phase-retrieval method are combined, which can overcome the influence of relative motion between the
imaging system and the object to imaging resolution in conventional imaging system.

For FGD, a single pointlike detector far from the object in the test path
can receive the global information of the object. However, the spectrum
center of FGD will move if we change the position $x_t$ of the single pointlike
detector, which can be explained by Eq. (7) because the center of the object's FGD is
related to the detection position $x_t$ and the result has also been demonstrated
in Ref. \cite{Liu}. Different from the case of
changing the detection position $x_t$, changing the object's center position
has no effect on FGD. As a result, the technique of FGD can overcome the influence
of relative motion between the imaging system and the object to the imaging resolution and the object's imaging with high spatial
resolution can be always reconstructed by phase-retrieval method from the FGD patterns, which
provides a brand-new approach of motion deblurring and is very useful to high-precision imaging systems and imaging of moving target.

In conclusion, we have given a proposal to improve the imaging
resolution of shaking object using ghost imaging technique. Both analytical and experimental results have shown that the object's
shaking has no effect on the imaging resolution of FGD and the object's imaging with high spatial
resolution can be always reconstructed from the FGD patterns. This
technique has a great application in practice to overcome motion blur.

The work was supported by the Hi-Tech Research and Development
Program of China under Grant Project No. 2011AA120101,
No. 2011AA120102.

\end{document}